# A Novel Approach for Implementing Steganography with Computing Power Obtained by Combining CUDA and MATLAB

Samir B. Patel[1], Shrikant N. Pradhan[2] and Saumitra U. Ambegaokar[3]

Computer Science and Engineering Department, Institute of Technology, Nirma University, SG-Highway, Ahmedabad, Gujarat, India.

[1] samir.patel@nirmauni.ac.in, [2] snpradhan@nirmauni.ac.in, [3] s.u.ambegaokar@gmail.com

*Abstract* --With the current development of multiprocessor systems, strive for computing data on such processor have also increased exponentially. If the multi core processors are not fully utilized, then even though we have the computing power the speed is not available to the end users for their respective applications. In accordance to this, the users or application designers also have to design newer applications taking care of the computing infrastructure available within. Our approach is to use the CUDA (Compute Unified Device Architecture) as backend and MATLAB as the front end to design an application for implementing steganography. Steganography is the term used for hiding information in the cover object like Image, Audio or Video data. As the computing required for multimedia data is much more than the text information, we have been successful in implementing image Steganography with the help of technology for the next generation.

## I. INTRODUCTION

The word steganography literally means covered writing as derived from Greek. It includes a vast array of methods of secret communications that conceal the very existence of the message. Among these methods are invisible inks, microdots, character arrangement (other than the cryptographic methods of permutation and substitution), digital signatures, covert channels and spread-spectrum communications. Steganography is the art of concealing the existence of information within seemingly innocuous carriers [1].

Steganography can be viewed as akin to cryptography. Both have been used throughout recorded history as means to protect information. At times these two technologies seem to converge while the objectives of the two differ. Cryptographic techniques "scramble" messages so if intercepted, the messages cannot be understood. Steganography, in an essence, "camouflages" a message to hide its existence and make it seem "invisible" thus concealing the fact that a message is being sent altogether. An encrypted message may draw suspicion while an invisible message will not.

Steganography has its place in security. It is not intended to replace cryptography but supplement it. Hiding a message with steganography methods reduces the chance of a message being detected. However, if that message is also encrypted, if discovered, it must also be cracked (yet another layer of protection). A transformation based digital watermarking techniques could be found in [2][3].

The tool which is prepared using MATLAB and CUDA is designed, to achieve steganography in image/video using LSB (Least Significant Bit) technique. In this technique, we modify the least significant bit positions of the image considering that the perceptual quality of the image is not sacrificed. Section 2 gives the introduction to NVIDIA-CUDA, section 3 provides basic architecture, section 4 provides the tools used for compiling, section 5 provides the complete road map for the work done, Section 6 and 7 discusses about the embedding and extraction process followed by the experimental results in section 8.

## II. INTRODUCTION TO NVIDIA CUDA

NVIDIA CUDA is a general purpose parallel computing architecture that leverages the parallel compute engine in NVIDIA graphics processing units (GPUs) to solve many complex computational problems in a fraction of the time required on a CPU. It includes the CUDA Instruction Set Architecture (ISA) and the parallel compute engine in the GPU. To program to the CUDA architecture, developers can, today, use C, one of the most widely used high-level programming languages, which can then be run at great performance on a CUDA enabled processor.

## III. ARCHITECTURE

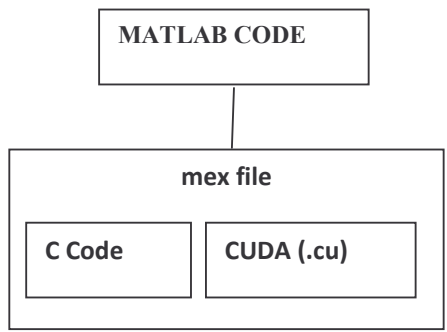

Figure 1. Basic Architecture

The basic architecture of the integration of Matlab with CUDA is shown in Figure1. The steganography tool is built by combining Matlab – which is used for reading image /video frames – with NVIDIA CUDA, for the LSB embedding of data. NVIDIA CUDA itself uses the C Compiler to compile the C portion of the CUDA file and nvcc (NVIDIA CUDA COMPILER) for compiling CUDA portion respectively. The CUDA file, along with a gateway file is compiled into a mex file (mexw32 or mexw64 on windows platform), which can be used from the MATLAB portion. A complete model under implementation is shown in Figure2 and CUDA gateway is shown in Figure 3

The following things are included in the interface:

- A CUDA wrapper – A *.cpp file – containing a mexw32 function, compiled to obtain a MATLAB mex file (similar to a .dll file).
- A CUDA gateway function – in *.cu file – this is the function called from the wrapper.
- A CUDA function – this is where the computation is actually called.

## IV. TOOLS USED FOR COMPILING

The following compilers are used:
i. CL compiler of Visual Studio 2005
ii. Nvcc of NVIDIA CUDA 2.1
iii. MEX compiler of MATLAB

To compile the *.cu file, the following function is used and is shown in Figure4.

The first line in the function: "!"%VS80COMNTOOLS%vsvars32.bat" & nvcc -c -deviceemu cudalsb.cu" is used to compile the file "cudalsb.cu" and create an object file. The C portion is compiled using Visual C++ compiler and the cuda portion with nvcc – Nvidia C Compiler.

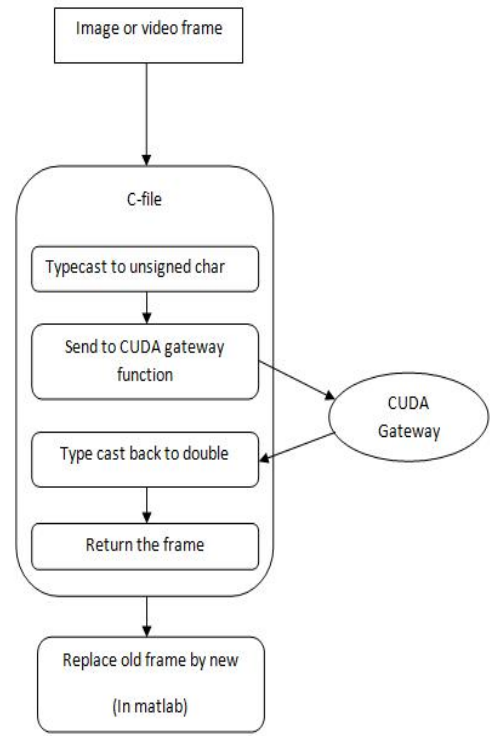

Figure 2. Complete model under implementation

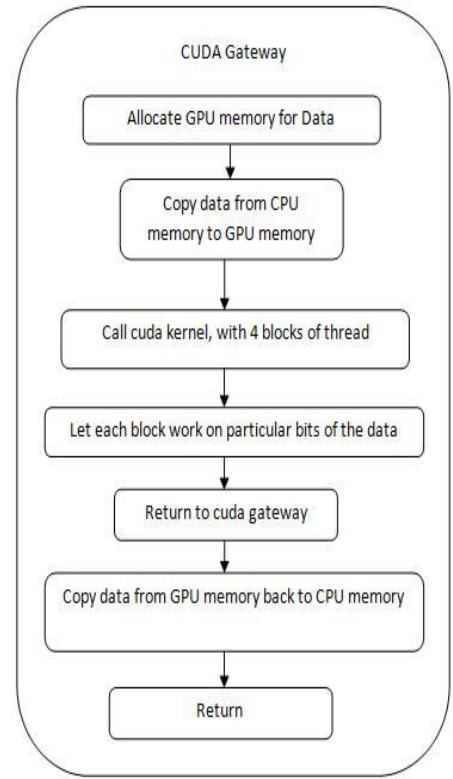

Figure 3. Flow inside the CUDA Gateway Function

```
function my_compile(varargin)
!"%VS80COMNTOOLS%vsvars32.bat" &
nvcc -c -deviceemu cudalsb.cu

n=getenv('CUDA_LIB_PATH');
if n(1)=='', n=n(2:end); end, if n(end)=='',
n=n(1:end-1);end

mex(['-L' n],'-lcudart','cudalsb.obj',varargin{:});
```

Figure 4. Function to compile

The 2nd and 3rd line are used to get the cuda library path. The last line is a call to mex, in order to create the mexw32 file.

['-L' n] – include directory with path given by string n.

cudalsb.obj is the object file to be included. varargin is the input to the my_compile function, which is the .cpp CUDA wrapper. At the execution of this cudalsb.obj and cudalsb.mexw32 are created.

## V. ALGORITHMIC EXPLANATION

The main MATLAB function calls the CUDA wrapper, which is a mexw32 file. The argument to this is a row of the frame along with the array containing the data to be embedded. The complete model including the flow inside the mex file is shown in Figure 3.

This function (which accepts only double arrays), typecasts the double array to an unsigned char type, so that bitwise operations can be done. Then, we pass the same to the CUDA gateway function. The CUDA gateway then allocates proper memory in the GPU. Then it copies the data to it. Now the cuda function is called, which performs appropriate bitwise operations in it and then returns to the gateway function.

Here the results are copied back to the main memory and the function returns to the cuda wrapper. In the wrapper, it is type-casted back to double and then returns to MATLAB function. The complete flow inside the CUDA function is shown in Figure 3 and Figure 4.

## VI. EMBEDDING PROCESS

The algorithm for embedding an array inside the image frame is shown below with the following parameters:
A as row of the image, B as array of data to be embedded, I as the block id of thread, j as the thread id and L as length of the data

1. Read an Image.
2. Separate any one plane (e.g. R plane).
3. Calculate the size of data to be embedded.
4. Based upon the size, divide data to be embedded among every row of image

5. For each row of image A, take an appropriately sized array of data B such that:

    size(B)*4 <= size(A)

6. Create 4 blocks of threads, each having 'n' threads, where

    n = min(32, length(data))

7. For each block with block id 'i' perform the following:
    a. Get thread id 'j'
    b. Get input data y = B[j]
    c. Mask y based upon the value of 'i'

        If i = 0 then mask with 0000 0011
        If i = 1 then mask with 0000 1100
        If i = 2 then mask with 0011 0000
        If i = 3 then mask with 110000 00

    d. Now right shift y bitwise by 2*i bits
    e. Obtain image pixel x = A[L*i + j]
    f. Mask x with 1111 1100 and logically OR it with y.
    g. Store this in resultant image

Repeat this process for the entire data to be embedded, passing a separate row of the image along with different data portion each time.

## VII. EXTRACTION PROCESS

As the process is symmetric the extraction process is the reverse of what is happening during the embedding process

A as row of the image, N as length of data embedded, i as the block id of thread, j as the thread id and L as length of the data

1. Read an Image.
2. Separate any one plane (e.g. R plane).
3. Based upon the size, divide data to be extracted among every row of image
4. For each row of image A, take an appropriately sized array of data B such that:

    size(B) = min(N, size(A)/4)

    N = N – size(B)

    Initialize B to zeroes.

5. Create 4 blocks of threads, each having 'n' threads, where

    n = min(32, length(data))

6. For each block with block id 'i' do the following:
    a. Get thread id 'j'
    b. Get input data y = A[L*i + j]
    c. Mask y with a mask 0000 0011
        y = y AND 0000 0011
    d. Now left shift y bitwise by 2*i bits

  e. Let x = B[j]

  f. Logically OR x with y.

  g. Store this back in B

Repeat the process for the entire data to be extracted, passing a separate row of the image each time.

## VIII.  EXPERIMENTAL RESULTS

**Direct Use of MEX functions:**

Figure 5 shows the original image used as the cover object for embedding, Figure 6 shows the image to be embedded and the Figure 7 shows the resultant image obtained after embedding and Figure 8 shows the recovered image. Likewise Figure 9 shows the stego image and Figure 10 shows the extracted image from Figure 9.

D: row of image 1024x1, A = 7x8 matrix (small image)

As we can send only a 1-D array, we send A(1:56) which gives a column major form.

>> E = cudalsb(D, A(1:56));

The output obtained is a column to an input of a row of the image.

So, before further operations, transpose of E is to be found.

>> E = transpose(E);

NOTE: If we do not take the transpose, then segmentation fault arises.

>> F = extract(E, 56);

Now the 7x8 array/frame is in form of row, so we have to get back the frame form.

To extract, we use a function which converts the column major form into a frame.

>> G = getimage(F, 7, 8);

>> imshow(uint8(G))

For the 7x8 image embedded, if we consider an 8-bit image, then we get a PSNR value of  51.6715.

For a 24-bit image, PSNR ratio increases to 56.4427

## IX.  CONCLUSION

NVIDIA CUDA is an emerging technology on multiple processors, we have been successful in implementing the steganographic application on the NVIDIA CUDA using MATLAB as the front end and CUDA as the back end. The outcome of the implementations shows that we have ended up developing a tool or utility in MATLAB which could be utilized for steganograhic application.

Future work will include the utilization of DCT and DWT based transformation techniques for performing digital watermarking using CUDA and MATLAB based implementation.

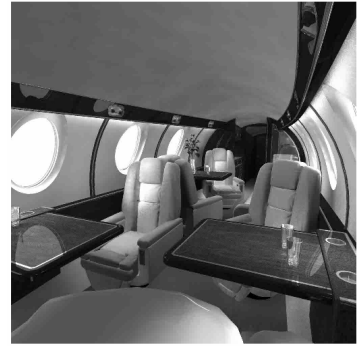

Figure 5. Original Cover Image (Aeroplane)

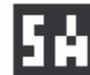

Figure 6. Image to be Embedded in the Cover Image

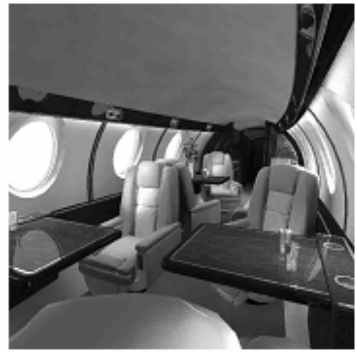

Figure 7. Cover Image obtained after doing embedding

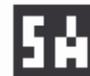

Figure 8. Recovered image.

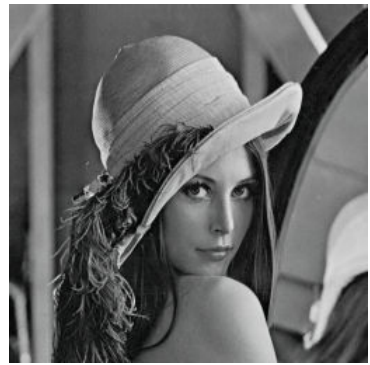

Figure 9. Cover Image obtained after doing embedding

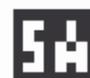

Figure 10. Recovered image.

## ACKNOWLEDGEMENT

We would like to thank all the members of NIRMA UNIVERSITY for providing continuous support and inspiration.

## AUTHORS PROFILE

**First Author:** Prof. Samir B. Patel is born in Ahmedabad, 26/07/1975 , He obtained his degree BE computer Engineering from LD college of Engineering, Ahmedabad, Gujarat, India in the year 1998, He did his M.E computer Engineering from SP University, Vallabh Vidhyanagar, Gujarat, India. His area of interest involved processing multimedia data, data Security, parallel computing. He is currently working at the post of Sr. Associate professor at CSE Department, Institute of Technology, Nirma University, Ahmedabad, Gujarat, India.

**Second Author:** Dr. Shrikant N. Pradhan is Ph.D. and he is having a vast experience of research. He was associated with Physical Research Laboratory (INDIA) for more than 25 years. He is currently Head of the M. Tech Programme at Institute of Technology, CSE Department, Nirma University., Ahmedabad, Gujarat, India. His area of interest involves signal processing, Embedded systems, Multimedia, Data Security and many more.

**Third Author:** Mr. Saumitra U. Ambegaokar is currently a student of B. Tech, Institute of Technology, CSE Department, Nirma University, Ahmedabad, Gujarat, India.